%% file: ms.tex
\documentclass[12pt,preprint]{aastex}

\newcommand{\Msun}{\mbox{$M_\odot$}\xspace}
\newcommand{\msun}{\Msun}
\def\lapp{\ifmmode\stackrel{<}{_{\sim}}\else$\stackrel{<}{_{\sim}}$\fi}
\def\gapp{\ifmmode\stackrel{>}{_{\sim}}\else$\stackrel{>}{_{\sim}}$\fi}
\def\deg{\ifmmode^\circ\else$^\circ$\xspace\fi}

\usepackage{xspace}
\usepackage[usenames, dvipsnames]{color}
\usepackage{soul}
\usepackage{longtable}
\usepackage{dcolumn}
\usepackage{multirow}
\newcolumntype{d}[1]{D{.}{.}{#1}}
\newcolumntype{X}[1]{D{-}{-}{#1}}

\newcommand{\presto}{\texttt{PRESTO}\xspace}

\newcommand{\rfifind}{\texttt{rfifind}\xspace}

\newcommand{\prepfold}{\texttt{prepfold}\xspace}

\newcommand{\tempo}{\texttt{TEMPO}\xspace}

\newcommand{\psrchive}{\texttt{psrchive}\xspace}
\newcommand{\pat}{\texttt{pat}\xspace}
\newcommand{\snr}{$S/N$\xspace}

\newcommand{\sigfourier}{\ifmmode\sigma_F\else$\sigma_F$\fi\xspace}
\newcommand{\snrtime}{\ifmmode\left ( S/N \right )_T\else$\left( S/N \right )_T$\fi\xspace}

\newcommand{\dmunit}{$\mathrm{pc\,cm^{-3}}$\xspace}

\newcommand{\sigmasp}{\ifmmode\sigma_\mathrm{block}\else$\sigma_\mathrm{block}$\fi\xspace}
\newcommand{\sigmaloc}{\ifmmode\sigma_\mathrm{loc}\else$\sigma_\mathrm{loc}$\fi\xspace}

\newcommand{\sect}{\S}

\newcommand{\psr}{PSR~J1913+1102\xspace}

\newcommand{\removed}[1]{}

\slugcomment{\input{./githash.tex}}

\begin{document}

\title{Einstein@Home discovery of a Double-Neutron Star Binary in the PALFA Survey}

\author{
P.~Lazarus\altaffilmark{1},
P.~C.~C.~Freire\altaffilmark{1},
B.~Allen\altaffilmark{2,3,4},
S.~Bogdanov\altaffilmark{5},
A.~Brazier\altaffilmark{6,7},
F.~Camilo\altaffilmark{8},
F.~Cardoso\altaffilmark{9},
S.~Chatterjee\altaffilmark{6},
J.~M.~Cordes\altaffilmark{6},
F.~Crawford\altaffilmark{10},
J.~S.~Deneva\altaffilmark{11},
R.~Ferdman\altaffilmark{12,13},
J.~W.~T.~Hessels\altaffilmark{14,15},
F.~A.~Jenet\altaffilmark{16}, 
C.~Karako-Argaman\altaffilmark{12,13},
V.~M.~Kaspi\altaffilmark{12,13},
B.~Knispel\altaffilmark{2,3},  
R.~Lynch\altaffilmark{17},
J.~van~Leeuwen\altaffilmark{14,15}, 
E.~Madsen\altaffilmark{12,13},
M.~A.~McLaughlin\altaffilmark{9},
C.~Patel\altaffilmark{12,13},
S.~M.~Ransom\altaffilmark{17},
P.~Scholz\altaffilmark{12,13},
A.~Seymour\altaffilmark{18},
X.~Siemens\altaffilmark{4}, 
L.~G.~Spitler\altaffilmark{1},
I.~H.~Stairs\altaffilmark{19,13}, 
K.~Stovall\altaffilmark{20}, 
J.~Swiggum\altaffilmark{8},
A.~Venkataraman\altaffilmark{18}
W.~W.~Zhu\altaffilmark{1},
}

\altaffiltext{1}{Max-Planck-Institut f\"ur Radioastronomie, Auf dem H\"ugel 69, 53121 
                 Bonn, Germany; pfreire@mpifr-bonn.mpg.de}
\altaffiltext{2}{Max-Planck-Institut f\"ur Gravitationsphysik, D-30167 Hannover, Germany}
\altaffiltext{3}{Leibniz Universit{\"a}t Hannover, D-30167 Hannover, Germany}
\altaffiltext{4}{Physics Dept., Univ. of Wisconsin - Milwaukee, Milwaukee WI 53211, USA}
\altaffiltext{5}{Columbia Astrophysics Laboratory, Columbia Univ., New York, NY 10027, USA} 
\altaffiltext{6}{Dept. of Astronomy, Cornell Univ., Ithaca, NY 14853, USA} 
\altaffiltext{7}{Center for Advanced Computing, Cornell Univ., Ithaca, NY 14853, USA}
\altaffiltext{8}{SKA South Africa, Pinelands, 7405, South Africa}
\altaffiltext{9}{Dept.~of Physics, West Virginia Univ., Morgantown, WV 26506, USA} 
\altaffiltext{10}{Dept.~of Physics and Astronomy, Franklin and Marshall College, Lancaster, 
                  PA 17604-3003, USA} 
\altaffiltext{11}{National Research Council, resident at the Naval Research Laboratory,
                  Washington, DC 20375}
\altaffiltext{12}{Dept.~of Physics, McGill Univ., Montreal, QC H3A 2T8, Canada}
\altaffiltext{13}{McGill Space Institute, McGill Univ., Montreal, QC H3A 2T8, Canada}
\altaffiltext{14}{ASTRON, the Netherlands Institute for Radio Astronomy, Postbus 2, 7990 AA, 
                  Dwingeloo, The Netherlands} 
\altaffiltext{15}{Anton Pannekoek Institute for Astronomy, Univ. of Amsterdam, Science 
                 Park 904, 1098 XH Amsterdam, The Netherlands}
\altaffiltext{16}{Center for Gravitational Wave Astronomy, Univ.~of Texas - Brownsville, 
                  TX 78520, USA} 
\altaffiltext{17}{NRAO, Charlottesville, VA 22903, USA} 
\altaffiltext{18}{Arecibo Observatory, HC3 Box 53995, Arecibo, PR 00612}
\altaffiltext{19}{Dept.~of Physics and Astronomy, Univ.~of British Columbia, Vancouver, 
                  BC V6T 1Z1, Canada} 
\altaffiltext{20}{Dept.~of Physics and Astronomy, Univ.~of New Mexico, NM 87131, USA}


\begin{abstract} We report here the Einstein@Home discovery of \psr, a 27.3-ms
pulsar found in data from the ongoing Arecibo PALFA pulsar survey. The pulsar is
in a 4.95-hr double neutron star (DNS) system with an eccentricity of 0.089.
 From radio timing with the Arecibo 305-m telescope, we measure the
rate of advance of periastron to be $\dot{\omega} = 5.632(18)$\,\deg/yr.
Assuming general relativity accurately models the orbital motion, this
corresponds to a total system mass of $M_\mathrm{tot} =
2.875(14)$\,\msun, similar to the mass of the most massive DNS known to date, B1913+16,
but with a much smaller eccentricity. The small eccentricity indicates
that the second-formed neutron star (the companion of PSR~J1913+1102) was
born in a supernova with a very small associated kick and mass loss.
In that case this companion is likely, by analogy with
other systems, to be a light ($\sim$1.2\,\msun)
neutron star; the system would then be highly asymmetric. A search
for radio pulsations from the companion yielded no plausible
detections, so we can't yet confirm this mass asymmetry.
By the end of 2016, timing observations should permit
the detection of two additional post-Keplerian parameters: the Einstein
delay ($\gamma$), which will enable precise mass measurements and a
verification of the possible mass asymmetry
of the system, and the orbital decay due to the emission of gravitational
waves ($\dot{P}_b$), which will allow another test of the radiative
properties of gravity. The latter effect will cause the system to
coalesce in $\sim$0.5\,Gyr.
\end{abstract}


\section{Introduction}
\label{sec:Introduction}

Studies of pulsars in double neutron star (DNS) binary systems provide exquisite
tests of relativistic gravity \citep[e.g.][]{ksm+06,wh16} and binary stellar
evolution \citep[e.g.][]{fsk+13}; they also allowed the first and still the most
precise neutron star (NS) mass measurements \citep{tw82,ksm+06}. Historically,
NS mass measurements were important to constrain the equation of state of
ultra-dense matter \citep[e.g.][]{lp04}. DNS orbits are observed to be slowly
contracting due to the emission of gravitational waves (GWs), thus demonstrating
their existence \citep[e.g.][]{tw82}; studies of the DNS population can be used
to inform the expected rate of Advanced LIGO detection NS-NS mergers
\citep{aaa+10}.

Pulsars in DNS systems are quite rare: out of $\sim$2500 known
radio pulsars, only 14 are known to be in DNS systems\footnote{There is still
some debate on the nature of the companion of two of these pulsars,
PSRs~J1906+0746 and J1807$-$2500B.}, including two in the ``double pulsar''
system, J0737$-$3039\footnote{In this work, we only use the prefix
PSR when referring to individual pulsars, in the latter system
PSR J0737$-$3039A and B. For the systems, we use the prefix DNS, unless we
refer to them specifically as a binary system.}. See \citet{msf+15} for a
recent summary of pulsars in DNS systems. 

DNS systems are typically found in large-scale, un-targeted radio pulsar surveys
such as the on-going Pulsar-ALFA (PALFA) survey with the 305-m William E. Gordon
telescope of the Arecibo Observatory in Puerto Rico \citep{cfl+06}. PALFA
observations are focused on the two regions of the Galactic plane visible from
Arecibo ($32\deg \lapp \ell \lapp 77\deg$ and $168\deg \lapp \ell \lapp 214\deg$
within $|b| < 5\deg$) and are conducted at 1.4\,GHz using the 7-beam Arecibo
L-Band Feed Array (ALFA) receiver and the Mock spectrometers (for a recent
overview of this survey, its observing set-up and its \presto-based data
reduction pipeline\footnote{\presto can be found at
https://github.com/scottransom/presto and the PALFA survey \presto-based
pipeline can be found at https://github.com/plazar/pipeline2.0.}, see
Lazarus et al. 2015\nocite{lbh+15}). In addition to the \presto-based pipeline, all PALFA survey
observations are analyzed using the Einstein@Home pulsar search pipeline
\citep[for details see][]{akc+13}. The relatively short, 268-s pointings of this
survey allows for the discovery of highly accelerated pulsars in compact binary
systems using relatively low computational cost ``acceleration'' searches. A
prime example is PSR~J1906+0746, the first DNS system found in the PALFA survey
and second most relativistic DNS system currently known \citep{lsf+06,lks+15}:
this system was detected with the ``quicklook'' pipeline without any
acceleration searches.

In this paper we present the discovery and early follow-up of \psr, a
member of a new DNS system. The paper is organized as follows:
The discovery and follow-up observations
are summarized in \sect\ref{sec:observations}. \sect\ref{sec:analysis}
describes the analysis and results from timing as well as a search for
radio pulsations from the companion of \psr. The results of these
analyses are discussed in \sect\ref{sec:discussion}, this includes
a discussion on the nature, eccentricity, total mass and possible mass
asymmetry of this new system. Prospects for future
timing observations of this binary system are presented in
\sect\ref{sec:prospects} before the paper is concluded in
\sect\ref{sec:conclusion}.

\section{Discovery and follow-up}
\label{sec:observations}

\psr was discovered by the Einstein@Home pipeline in a PALFA observation taken
on 2010 September 26 (all dates and times are in UT).
The pulsar has a spin period of 27.3 ms and a dispersion
measure (DM) of 339\,\dmunit. For the pulse profile, the reduced $\chi^2$
was 3.3, which corresponds to a detection significance of 8.5 $\sigma$
for the best acceleration ($-54 \pm 14 \rm \, m \, s^{-2}$). For zero 
acceleration, the reduced $\chi^2$ is only 1.4, which corresponds to a
significance of 5.5 $\sigma$.
This implies that this pulsar is in a highly accelerated binary system and that
it would not have been discovered without the acceleration search algorithms.

Following its discovery, \psr was observed at the Arecibo Observatory using the
ALFA receiver and Mock spectrometers using an observing set-up identical to
PALFA survey observations \citep[see][for details]{lbh+15}. In short, the
322-MHz observing band centered at 1375\,MHz is divided into two overlapping
sub-bands. Each 172\,MHz sub-band is divided into 512 channels and is sampled
every $\sim$65.5\,$\mu$s. In this observing mode, the two linear polarizations
are summed and only the total intensity is recorded. In total, 11 observations
were conducted in this mode between 2012 May 25 and 2013 September 20.
Integration times ranged between 110~and 1200\,s.

The variations of the measured period of the pulsar in these
observations established that \psr is in a 4.95-hr binary orbit with an
eccentricity of $e\simeq0.09$. The mass function, $M_f \simeq 0.136$\,\msun,
yields, assuming a pulsar mass $M_p=1.35$\,\msun, a companion mass $M_c \, > \,
0.878$\,\msun.

\begin{figure}
\includegraphics[width=3.2in]{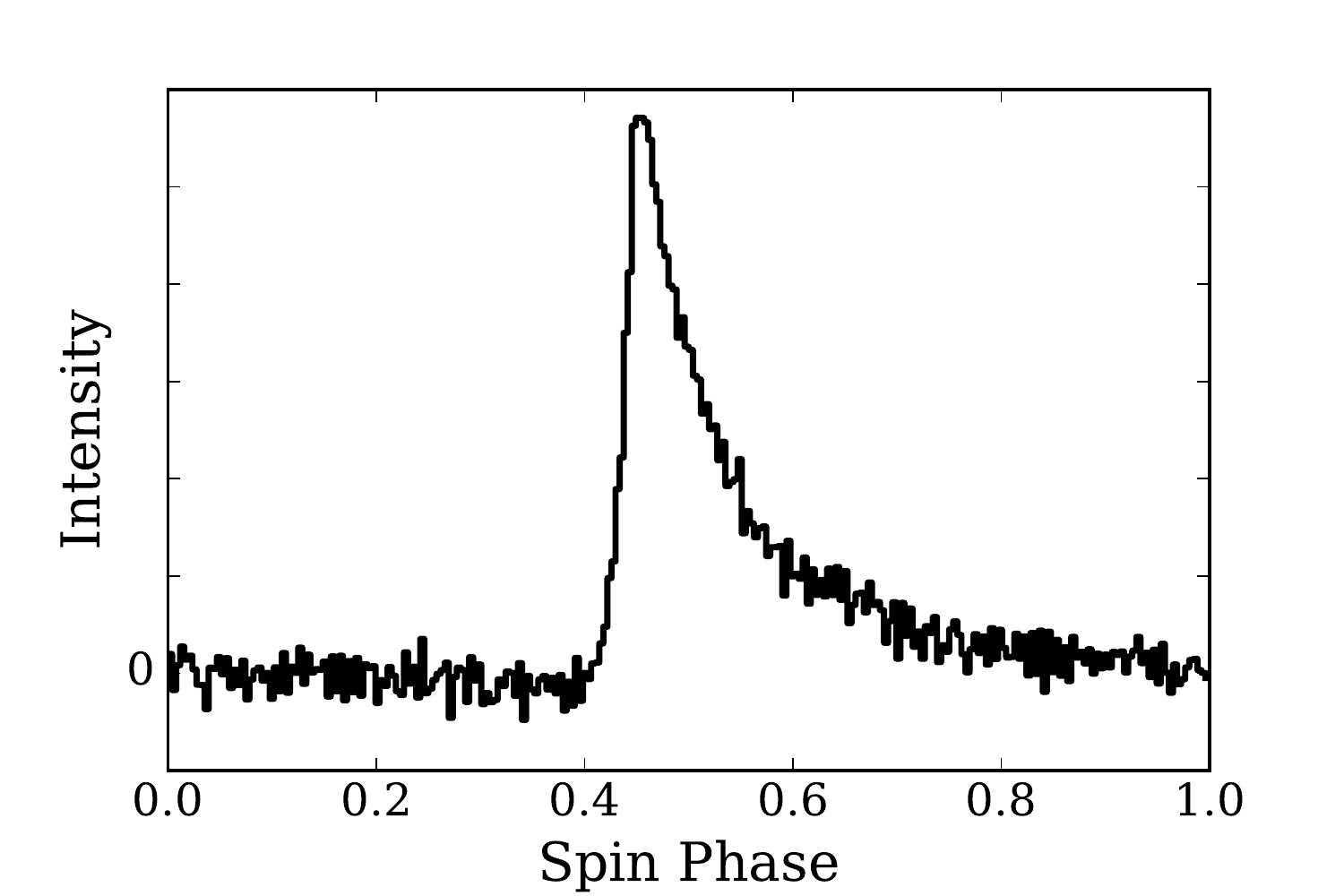}
\\
\includegraphics[width=3.2in]{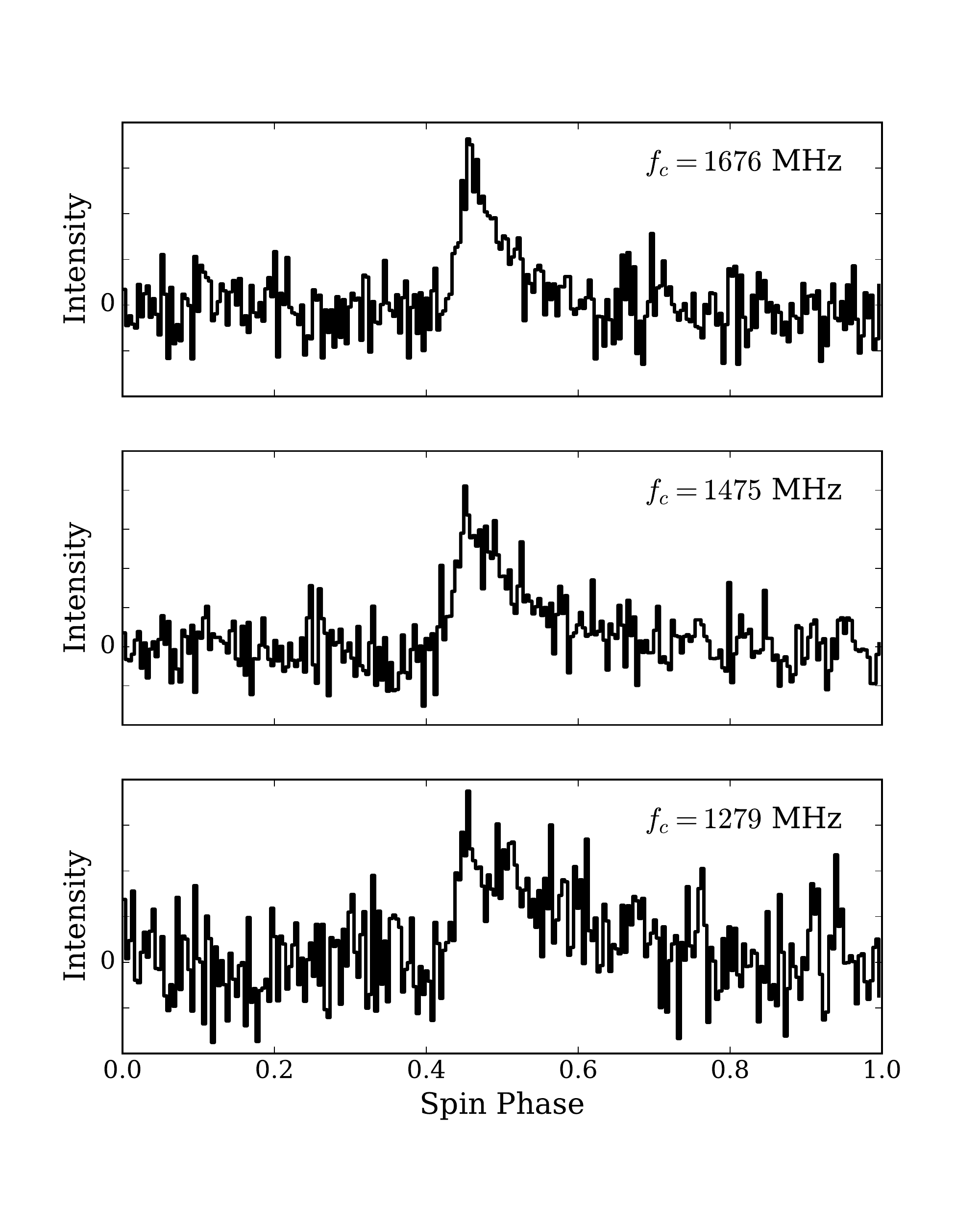}
\caption{{\em Top}:A high-\snr ($\snr=98.7$) profile of \psr at L-band (centered at
1380 MHz) created by phase-aligning
and summing all 36 incoherent PUPPI observations. A total of 5.3\,hr of data
were summed.  Details of the observing set-up used for the observations
included in the profile can be found in \sect~\ref{sec:observations} and
Table~\ref{tab:obs_summary}. The pulse profile exhibits exponential
decay after the main pulse, this is likely caused by interstellar
scattering, a common occurrence for pulsars at high DMs.
{\em Bottom}: Pulse profile from one of the longest observations
at three separate sub-bands. We can see here that the exponential decay
becomes much longer at lower frequencies, as expected from interstellar
scattering.
\label{fig:profile}}
\end{figure}

Starting on 2012 November 10, \psr was also observed with the L-wide receiver at
the Arecibo Observatory using the Puerto Rico Ultimate Pulsar Processor
Instrument (PUPPI) backend in ``incoherent'' mode. These L-wide PUPPI
observations contain 600\,MHz of usable bandwidth centered at 1380 MHz
uniformly sub-divided into
2048 channels.  \psr was observed with this set-up 36 times before 2014 January
1 (the cut-off for this work). All 36 PUPPI "incoherent" mode observations
were phase-aligned and integrated together to form a high-\snr ($\snr=98.7$)
L-band profile of \psr. The resulting profile is shown in the top plot of
Fig.~\ref{fig:profile}. The average pulse profile has a half-power duty cycle
of $\sim \, 0.08$. It shows an exponential decay after the main
pulse, as if the signal is affected by interstellar scattering;
this is not too surprising given the high DM of the pulsar.
This is confirmed by the fact that the exponential decay timescale becomes
much longer at lower frequencies, as shown in the lower plot of Fig.~\ref{fig:profile}.

\begin{table}
\centering
\caption{Observation Summary
\label{tab:obs_summary}}
\begin{scriptsize}
\begin{tabular}{ccccccccc}
    \hline\hline\\[-2mm]
    & & &  & Center & Total & & & \\[1mm]
    Receiver & Data Recorder & Mode & Span  & Frequency & Bandwidth & No. of channels & No. Obs. & No. TOAs\\[1mm]
             &               &      & (MJD) & (MHz)     & (MHz) & & \\[1mm]
    \hline\\[-2mm]
    ALFA     & Mock (Low band) & Search & 56072--56555 & 1300.168 & 172.032 & 512 & 11 & 53 \\
    ALFA     & Mock (High band) & Search & 56072--56555 & 1450.168 & 172.032 & 512 & 11 & 55 \\
    L-Wide   & PUPPI & Search & 56241--56642 & 1440 & 800$^a$ & 2048 & 36 & 405 \\
    \hline \\[-2mm]
\end{tabular}
\raggedright
\\$^{a}$The L-wide receiver has a bandwidth of $\sim$600\,MHz, only that part of the band
is scientifically useful.\\
\end{scriptsize}
\end{table}

A summary of the observations included in this
paper is presented in Table~\ref{tab:obs_summary}.
All of these observations were conducted in search mode because
at the time the observations were taken we still lacked a phase-coherent timing
solution that would allow for coherent dedispersion and online folding.
The advantage of search data is that they can be dedispersed and folded
with improved
ephemerides at a later time; this allows for iterative improvement of the
ephemeris. They also enable the search for pulsations from the neutron star
companion (see \sect\ref{sec:companionsearch}), which is not possible with
timing data. The disadvantage is that these data have some dispersive smearing,
and were not properly flux and polarization calibrated.

Nevertheless, we can estimate an approximate flux density at 1.4 GHz
based on the radiometer equation \citep{dtws85} and the PUPPI pulse profile in Fig. 1.
Using the characteristics of the L-wide receiver
(a gain of 10.5 K/Jy, a bandwidth of 600 MHz, a system temperature
of $\sim\,$30 K\footnote{http://www.naic.edu/$\sim$astro/RXstatus/rcvrtabz.shtml},
to which we add 4 K of background sky temperature)
and the characteristics of the detection
(total observing time of 5.3 hours, pulse duty cycle of $\sim\, $8 \% and
signal-to-noise ratio of 98.7)
we derive a phase-averaged flux density at 1.4 GHz of $S_{1.4}\, \sim\, 0.02 \, \rm mJy$.
For the estimated distance of 7.6 kpc this represents a 1.4 GHz luminosity
of $L_{1.4}\, \sim\, 1.1 \, \rm mJy \, kpc^2$.
More rigorous estimates of flux density, spectral index, scattering times
and multi-frequency polarimetric properties of this pulsar will be published
elsewhere.

\section{Data Analysis and Results}
\label{sec:analysis}
\subsection{Timing Analysis}
\label{sec:timing}

The observations presented in Table~\ref{tab:obs_summary} were used to derive a
timing solution for \psr using standard pulsar timing techniques \citep[see][and
references therein]{lk04}. First, the data that were obviously affected by
radio frequency interference (RFI) were removed; the remaining data
were then incoherently dedispersed at DM\,=\,338.96\,\dmunit
and then folded at the topocentric period of the pulsar. Each of the two ALFA
sub-bands were reduced independently and the 600\,MHz L-wide band was divided
into three 200\,MHz sub-bands.

Three standard template profiles were created for the high-frequency ALFA band,
the low-frequency ALFA band, and the PUPPI band. These templates were
cross-correlated against the corresponding folded profiles to determine pulse
times-of-arrival (TOAs) using the ``FDM'' algorithm of \psrchive's \pat program.
The TOAs were fit to a model that accounts for spin,
astrometry and binary motion, including the relativistic rate of advance of
periastron, $\dot{\omega}$, which was found to be significant. The timing model
was fit using \tempo\footnote{http://tempo.sourceforge.net/}. In fitting the
data we used the NASA JPL DE421 Solar System ephemeris \citep{fwb09}, and the
NIST's UTC time scale\footnote{http://www.nist.gov/pml/div688/grp50/NISTUTC.cfm}. 

The TOAs were split into three sub-sets for the 1) low-frequency ALFA band, 2) 
high-frequency ALFA band, and 3) L-wide receiver. For each sub-set, the TOA 
uncertainties were scaled (using an ``EFAC'' parameter) such that the resulting 
reduced $\chi^2 = 1$ when the sub-set was fit independently of the others. The
multiplicative factors were found to be between 0.89 and 1.0.

The complete fit of all 513 TOAs had a reduced $\chi^2$ of 1.0 and a weighted
RMS of the timing residuals of 114\,$\mu$s. The fitted and derived timing
parameters are shown in Table~\ref{tab:timing}.

\begin{table}
\begin{footnotesize}
\caption{Fitted and derived parameters for \psr.}
\label{tab:timing}
\begin{tabular}{lc}
        \hline\\[-2mm]
        Parameter & Value$^{a}$\\
        \hline\\[-2mm]
        \multicolumn{2}{c}{\textit{General Information}} \\
        \hline\\[-2mm]
        MJD Range & 56072 -- 56642 \\
        Number of TOAs & 513 \\
        Weighted RMS of Timing Residuals ($\mu$s) & 114 \\
        Reduced-$\chi^2$ value$^{b}$ & 1.0 \\
        Reference MJD & 56357 \\
        Binary Model Used & DD \\
	Phase-averaged flux density at 1.4 GHz, $S_{1.4}$ (mJy) & $\sim$0.02 \\
	Pseudo-luminosity at 1.4 GHz, $L_{1.4}$ (mJy) & $\sim$1.1 \\
        \hline\\[-2mm]
        \multicolumn{2}{c}{\textit{Fitted Parameters}} \\
        \hline\\[-2mm]
        Right Ascension, $\alpha$ (J2000) & 19:13:29.0542(3) \\
        Declination, $\delta$ (J2000) & $+$11:02:05.741(9) \\
        Spin Frequency, $\nu$ (Hz) & 36.65016488379(2) \\
        Spin Frequency derivative, $\dot{\nu}$ ($\times 10^{-16}$ Hz/s) & $-$2.16(3) \\
        Dispersion Measure, DM (\dmunit) & 338.96(2) \\
        Projected Semi-Major Axis, $a\,\sin i$ (lt-s) & 1.754623(8) \\
        Orbital Period, $P_b$ (days) & 0.206252330(6) \\
        Time of Periastron Passage, $T_{0}$ (MJD) & 56241.029660(5) \\
        Orbital Eccentricity, $e$ & 0.08954(1) \\
        Longitude of Periastron, $\omega$ ($^{\circ}$) & 264.279(9) \\
        Rate of Advance of Periastron, $\dot{\omega}$ (\deg/yr) & 5.632(18) \\
        \hline\\[-2mm]
        \multicolumn{2}{c}{\textit{Derived Parameters}} \\
        \hline\\[-2mm]
        Spin Period, (ms) &  27.28500685251(2) \\
        Spin Period Derivative ($\times 10^{-19}$ s/s) & 1.61(2) \\
        Galactic longitude, $l$ ($^{\circ}$) & 45.25 \\
        Galactic latitude, $b$ ($^{\circ}$) & 0.19 \\
        Distance (NE2001, kpc) & 7.6 \\
        Characteristic Age, $\tau_c = P/(2\dot{P})$ (Gyr) & 2.7 \\
        Inferred Surface Magnetic & \multirow{2}{*}{2.1} \\
        Field Strength, $B_S$ ($\times 10^{9}$ G) & \\
        Spin-down Luminosity, $\dot{E}$ ($\times 10^{32}$ ergs/s) & 3.1 \\
        Mass Function, $f$ (\msun) & 0.136344(2) \\	
        Total Binary System Mass, $M_\mathrm{tot}$ (\msun)& 2.875(14) \\
        \hline\\[-2mm]
\end{tabular}
\end{footnotesize}
\raggedright
\\$^{a}$The numbers in parentheses are the 1-$\sigma$, \tempo-reported
uncertainties on the last digit.\\
$^{b}$The uncertainties of the two ALFA data sets and the L-wide data set were
individually scaled such that the reduced $\chi^{2}$ of the resulting residuals
are 1.
\end{table}

\subsection{Searching for the Companion of \psr}
\label{sec:companionsearch}

Given the large mass function for this system, it is likely that
the companion of \psr is another neutron star
(see Section~\ref{sec:discussion} for details). If so, then it could in principle
be an active radio pulsar, as observed in the J0737$-$3039 ``double pulsar" 
system. Motivated by
this possibility, we have searched for the presence of periodic signals in all
available observations with integration times longer than $T \gapp110$\,s.
This amounts to 11 observations with the central beam of ALFA with 110 $\lapp T
\lapp$ 1200\,s and 33 observations with the L-wide receiver with durations
between $\sim$300\,s and 1200\,s. In all, the observations searched span
1.5\,years.
Searching data sets with long time spans and covering all orbital
phases is important given that pulsars in DNS systems might be eclipsed and/or
precess into and out of the line of sight \citep[e.g.][]{bkk+08}.

The low- and high-frequency sub-bands of the Mock Spectrometer ALFA observations
were combined and excised of RFI using the methods described in \citet{lbh+15}.
Likewise, several often-corrupted frequency channels were zero-weighted in
the incoherent PUPPI observations.
RFI masks created with \presto's \rfifind were applied to the data
when dedispersing; after dedispersion we reduce the spectral resolution
by a factor of 16, this results in a total of 128 channels.
Both raw and zero-DM filtered \citep[see][]{ekl09}
dedispersed time series were produced and searched. Binarity was
removed from dedispersed time series using the technique described below. 

Most of the orbital parameters of the companion's binary motion are known
because the orbit of \psr has been precisely determined from our timing
analysis. The only unknown parameter is the projected semi-major axis of the
companion's orbit, $x_c = a_c \sin i$, where $a_c$ is the semi-major axis of
the companion's orbit and $i$ is the orbital inclination. The ratio of the size
of the pulsar's orbit and the companion's orbit is related to the unknown mass
ratio: $q=M_p/M_c=a_c/a_p$, where $M_p$ and $M_c$ are the pulsar and companion
masses, respectively.

Prior to searching for periodicities in the data, each observation was
dedispersed at the DM of \psr. The dedispersed time series were transformed to
candidate companion rest frames using a custom script that has been incorporated
into \presto. These were derived for 5000 evenly spaced trial values of
companion mass, $M_c \in [1.04-2.4]$\,\msun. Then, using the total mass of the
system $M_\mathrm{tot}$ (section \ref{sec:mass}), we calculate the pulsar
mass as $M_\mathrm{tot} - M_c$. From this we calculate $q$ and $x_c$.
With the rest frame established the dedispersed
time series were transformed by adding or removing samples as necessary to keep
each sample within 0.5 samples of its corrected value. When a sample is added to
the time series its value is set to the value of the preceding sample.

For each of the resulting 5000 time series, the Fast Fourier Transform (FFT) was
computed, normalized (including red noise suppression), and searched for
un-accelerated signals using 16-harmonic sums.  The output candidate lists were
sifted through to find promising signals to fold. Significant candidate signals
found at the same period in the same observation were grouped together.
Likewise, harmonically related candidates were grouped together with the
fundamental. The 20 most significant candidates in each observation were folded.
Each candidate was folded using the mass ratio at which it was most strongly
detected, as well as up to five other mass ratios uniformly distributed over the
range of ratios in which the signal was detected. All folding was done with
\presto's \prepfold. Each of the resulting diagnostic plots was inspected
manually. The most pulsar-like candidates were compared against a list of known
RFI signals from the PALFA survey.  Candidates with non-RFI-prone frequencies
were re-folded using full radio-frequency information. This procedure was
validated by applying it to an observation of the J0737$-$3039 system. Using the
ephemeris for PSR~J0737$-$3039A \citep{ksm+06} as a starting point, we were able
to detect PSR~J0737$-$3039B.

None of the candidates identified in the search for the companion of \psr was
consistent with coming from an astrophysical source. The minimum detectable flux
density of each observation was computed using the radiometer equation, with
the same parameters used in Section 2. 
Because the companion of \psr would likely be a slow, normal pulsar like
PSR~J0737$-$3039B, we did not include the broadening effects of the ISM (DM
smearing and scattering) and simply assumed a 5\,\% duty cycle. Furthermore,
for simplicity, we did not include the degradation of sensitivity due to red
noise found by \citet{lbh+15}. With these caveats, we derive
an upper limit for the 1.4 GHz flux density of about
$S_{1.4}\, < \, 12 \, \rm \mu Jy$ for most of the Mock observations, and
$S_{1.4}\, < \, 9 \, \rm \mu Jy$ for the longest the Mock observations.
For most of the L-wide/PUPPI observations, we obtain
 $S_{1.4}\, < \, 7 \, \rm \mu Jy$.

However, in practice these sensitivity limits are not reached in real surveys,
where, due to a variety of factors, there is always some sensitivity degradation,
especially for slow-spinning pulsars \citep{lbh+15}. Not knowing the spin period of the
companion of PSR~J1913+1102, it is not possible to estimate the degradation
factor precisely. If we assume a spin period of the order of a second, as in the
case of PSR~J0737$-$3039B, then for
DMs of 325 cm$^{-3}\,$pc, Lazarus et al. (2015) estimate a loss of sensitivity 
of $\sim$2. Doubling the flux density limit calculated above
would translate to a 1.4 GHz luminosity limit of
 $L_{1.4}\, < \, 0.8 \, \rm mJy \, kpc^2$. This would place
the companion among at the lowest 6 \% percentile of all pulsars in the ATNF
catalog \citep{mht+05} with reported flux densities at 1.4 GHz 
(\psr itself is in the lowest 8 \% in luminosity). 


\section{Discussion}
\label{sec:discussion}

\subsection{Nature of the companion}

Given the orbital parameters of \psr it is clear its companion is very massive,
at least 1.04 \msun (see section \ref{sec:mass}). Therefore, it could be either
a massive WD or another NS. For instance, in the case of PSR~J1141$-$6545,
a binary pulsar with very similar orbital parameters,
the companion to the pulsar is a massive WD \citep{abw+11} with a mass of
$\sim \, 1.0 \, M_{\odot}$
\citep{bbv08}, which is similar to our lower mass limit for the companion.

 However, the measured spin period derivative of \psr implies a
B-field of $2.1 \, \times \, 10^{9}$\,G and $\tau_c$ of 2.7\,Gyr. The characteristic age is
much larger, and the B-field is much smaller, than observed for PSR~J1141$-$6545
and the normal pulsar population in general. This implies that, unlike in the
case of PSR~J1141$-$6545, \psr was recycled by accretion of matter from the companion's
non-compact progenitor.
During this recycling, the orbit was very likely circularized, as normally
observed in all compact X-ray binaries. This means that something must have
induced the currently observed orbital eccentricity; the best candidate for this
would be the sudden mass loss and the kick that resulted from a second supernova
in the system. Thus, the companion is very likely to be another neutron star.
Its non-detection as a pulsar could mean that either it is no longer active as a
radio pulsar, or that its radio beam does not cross the Earth's position, or
alternatively that it is just a relatively faint pulsar. This situation
is similar to all but one of the known DNSs, where a single NS (in most cases
the first formed) is observed as a pulsar; the sole exception is,
of course, the ``double pulsar" system J0737$-$3039.

\subsection{The most massive DNS known}
\label{sec:mass}

With the data included in this paper, we have already detected the rate of
advance of periastron, $\dot{\omega}$.  If we assume this effect to be caused
purely by the effects of general relativity (GR) (a safe bet considering the
orbital stability of the system and the lack of observation of any variations
of the dispersion measure with orbital phase), then it depends, to first
post-Newtonian order, only on the total mass of the system $M_\mathrm{tot}$
and the well known Keplerian orbital parameters \citep{tw82}:

\begin{equation}
        \dot{\omega} = 3 T_{\odot}^{2/3} \left( \frac{P_{\mathrm{b}}}{2\pi } \right)^{-5/3} \frac{1}{1- e^2} M_\mathrm{tot}^{2/3}.
\end{equation}

\noindent $T_\odot$ is one solar mass expressed in units of time, $T_\odot
\equiv G \msun c^{-3} = 4.925490947$\,$\mu$s.
The $\dot{\omega}$ observed yields $M_\mathrm{tot} = 2.875 \pm 0.014$\,\msun,
making J1913+1102 significantly ($0.047 \pm 0.014\,$ \msun) more massive than
the B1913+16 system, until now most massive DNS
known \citep[$M_\mathrm{tot} = 2.828$\,\msun;][]{wh16}. 

\begin{figure}
\includegraphics[width=\columnwidth]{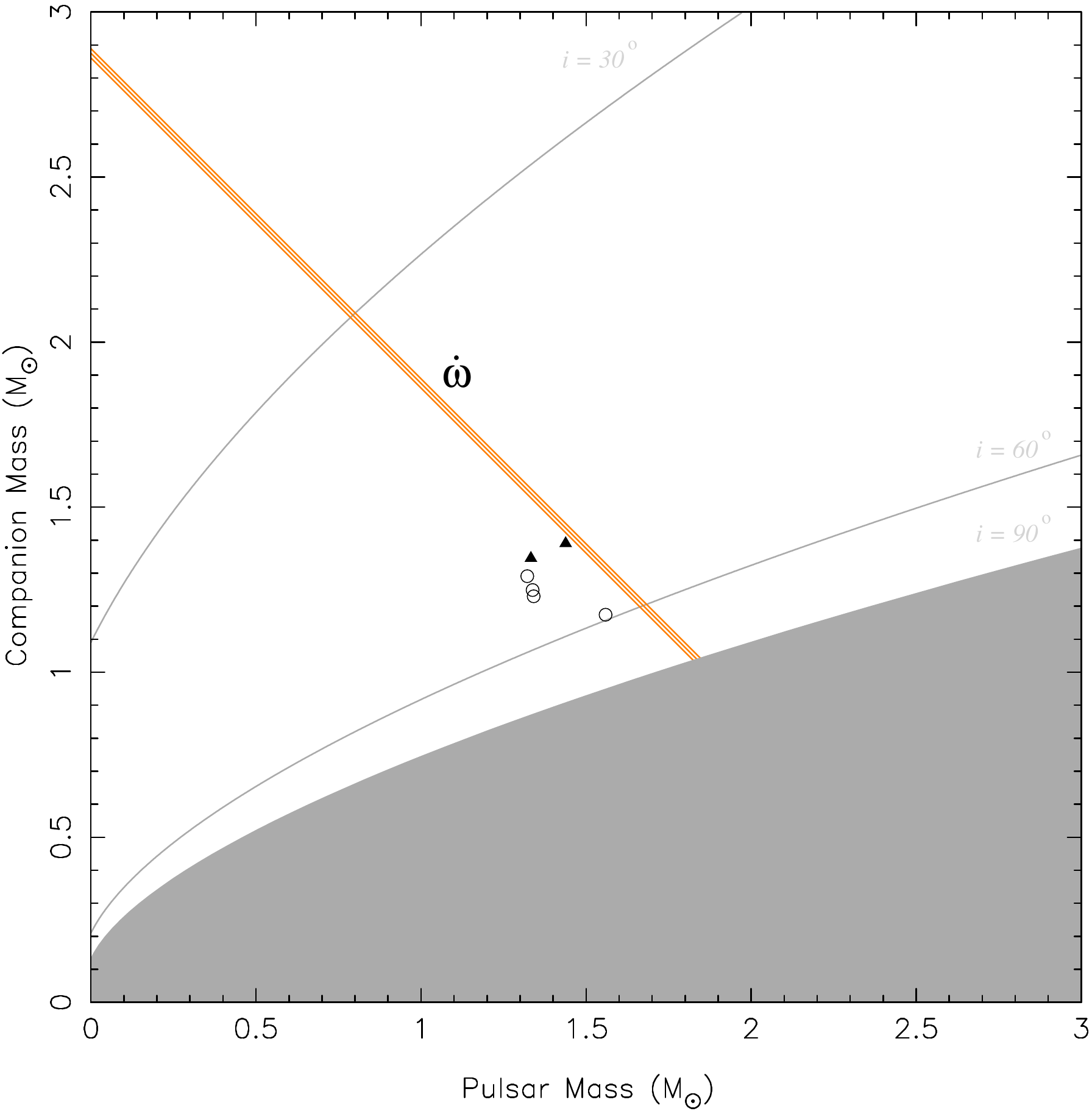}
\caption{Current mass constraints from the timing of PSR~J1913+1102.
The straight lines correspond to the nominal value and $\pm 1 \sigma$ uncertainty
of the rate of advance of periastron $\dot{\omega}$. The gray region is excluded
by the mass function of the system, and the gray lines indicate constant
orbital inclinations of $60^\circ$ (median) and $30^\circ$.
The symbols indicate the neutron star masses
for DNSs where these have been well measured, the mass of the first formed
neutron star is displayed along the x axis. The filled triangles indicate the
high-eccentricity DNSs, B1913+16 and B1534+12, the open circles
indicate the low-eccentricity DNSs like J1913+1102.\label{fig:mass_mass}}
\end{figure}

It is not yet possible to measure a second PK effect precisely enough to
estimate the individual NS masses. However, we can use the known mass function
to estimate an upper limit to the pulsar mass ($M_p \, <\, 1.84 \,\Msun$) and a
lower limit to the companion mass ($M_c \, > \, 1.04 \,\Msun$, see
Fig.~\ref{fig:mass_mass}).  These extreme values would occur for an orbital
inclination of 90\deg, i.e., if the system were being seen edge-on.

\subsection{The 0.09 clump in orbital eccentricity}

The J1913+1102 system is one of the six known DNSs in the Galactic  disk that
will coalesce in a time shorter than a Hubble time, the others being
J0737$-$3039 \citep{bdp+03,lbk+04,ksm+06},  J1906+0746 \citep{lks+15},
J1756$-$2251 \citep{fsk+14}, B1913+16 \citep{wh16} and B1534+12 \citep{fst14};
one of these, J1906+0746, is not securely identified as a DNS. Of these six
only J0737$-$3039 and J1906+0746 have smaller orbital periods than \psr.
These three systems have remarkably similar eccentricities: $0.0878$ for
J0737$-$3039, $e\,=\,0.0853$ for J1906+0746 and $e\, = \, 0.0896$ for
J1913+1102.

One question that naturally arises from this similarity is whether this is
merely a coincidence or whether it reflects a deeper similarity between these
three systems. One of the features of gravitational wave emission is that it
will not only cause a steady decrease in the orbital period of a system, but it will also
cause a decrease in the orbital eccentricity with time. This means that, for any
group of DNS systems, we will generally see a correlation between orbital period
and eccentricity.

\begin{figure}
\includegraphics[width=\columnwidth]{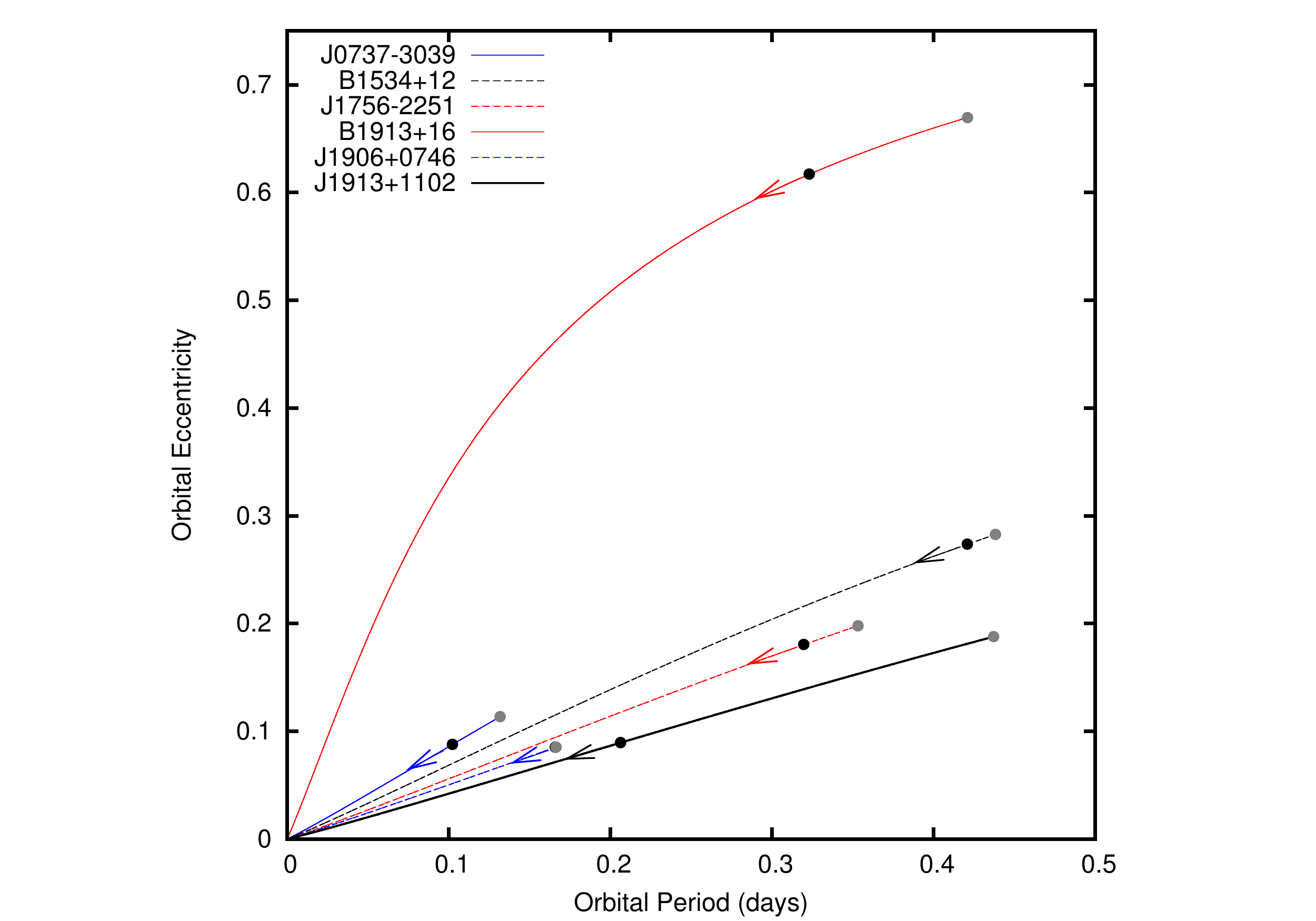}
\caption{Past and future orbital evolution of a set of 6 DNSs, which will coalesce within a Hubble time. Because of orbital energy
loss due to emission of gravitational waves, all systems are moving to the
lower left corner, with ever smaller orbital periods and orbital
eccentricities, reaching coalescence at the origin of the plot. The current
positions are given by black dots; our best estimates of the
positions right after the second SN are given by gray dots. \label{fig:ecc_pb}}
\end{figure}

In Fig.~\ref{fig:ecc_pb}, we see the past and future evolution of $e$ versus
$P_b$ for the six aforementioned DNSs, which have short coalescence times. All
systems are moving towards smaller $e$ and $P_b$, i.e., towards the lower left
corner.

Regarding the past evolution, we calculate the orbital evolution until a time that is
given by the best available proxy for the ``age" of the binary, by which
we mean the time elapsed since the second supernova. The positions for
those initial times are given by the gray-filled points.

In the case of J0737$-$3039 and J1906+0746 (interestingly, the two systems with
the shortest orbital periods), the second-formed NSs are observed as pulsars;
their $\tau_c$ allow reasonable constraints on the age of the systems. For
J0737$-$3039, the age of the system is either $\sim$80~or 180\,Myr
\citep{lfs+07}; for the sake of argument we use the lower estimate.  At that
time $e$ was about 0.114. For J1906+0746, the  young pulsar (the only one
detectable in that system) has $\tau_c \, \sim\,$100\,Kyr; this number is so
small that even if the true age of the system were larger by one order of
magnitude, there was just no time for significant change in the orbital
eccentricity. Therefore, its current eccentricity is very similar to
the orbital eccentricity right after the formation of the second NS.

For the other DNSs, including J1913+1102, we do not currently detect the young
pulsars, so we must rely on the $\tau_c$ of the recycled pulsars. In the case of
J0737$-$3039, PSR~J0737$-$3039A has $\tau_c\,=\,$200 Myr \citep{lbk+04}, which
is larger than the age estimated from the spin-down of PSR~J0737$-$3039B; the
same will likely be the case for the other systems. For \psr, $\tau_c \, = \,
2.7\,$Gyr; projecting the orbital parameters that far in the past we obtain
upper limits for $e$ and $P_b$ of 0.19 and 10.5\,hours respectively.  Not
knowing the $\tau_c$ of the second-formed NSs, we can't derive better
constraints on the birth parameters for J1913+1102 and most other DNSs (although
in the cases of B1534+12 and J1751$-$2251, they were likely similar to their
current parameters).

Nevertheless, the disparity between the initial orbital eccentricities of
J0737$-$3039 and J1906+0746 and the poorly constrained initial $e$ for
J1913+1102 do not positively suggest the existence of a tight clump of orbital
eccentricities at $\sim\,$0.09 for the DNSs with the shortest orbital periods.

Looking towards the future evolution of J1913+1102, the system will coalesce in
$\sim$0.5\,Gyr. These numbers are not yet very precise because we do not yet 
know the individual masses. In this calculation we use the values in the
discussion below ($M_p \sim\,$1.65\,\msun and $M_c \sim\,$1.24\,\msun); the
coalescence time does not change much for a more symmetric system.

\subsection{What do the low eccentricities mean?}

The observed DNS eccentricities nevertheless show that most of the compact
systems have low eccentricities, in marked contrast to
B1913+16.
After the discovery of J0737$-$3039, \citet{heu04} pointed out that in the
eccentric DNS systems (B1913+16 and B1534+12) the second-formed NSs have masses
of 1.389 and 1.346\,\msun and the systems have high proper motions, while for
the low-eccentricity systems (at the time J0737$-$3039, and now also including
J1756$-$2251 and J1906+0746) the second-formed NSs have masses in the range
1.23\,--\,1.29\,\msun, and at the time no detectable proper motions;
this is consistent
their location near the Galactic plane despite their large ages.
The masses observed in the J1141$-$6545 system are consistent with the latter scenario:
the pulsar formed {\em after} its heavy WD companion, but it has a relatively
low mass of $1.27\, M_{\odot}$, and the post-SN orbital eccentricity
of the system is 0.17.

All of this is believed to reflect the bimodal nature of the supernovae (SNe)
that formed the second NS,
with one group of SNe, those that form lighter NSs, having much smaller kicks
and associated mass loss. These are thought to be the result of ultra-stripped
SNe, which can either be electron capture or iron core-collapse SNe
\citep{tlm+13,tlp15}. 

Since then, the small kick of the second SN has been confirmed in studies of
two low-eccentricity DNSs: The proper motion of J0737$-$3039 has been
measured \citep{dbt09}; the tangential space velocity $v_t$ was been found to
be $\sim \, 10 \rm \, km \, s^{-1}$, much smaller
than observed for other pulsars and also small compared to the general population
of recycled pulsars (for J1756$-$2251 the limits are not so constraining,
$v_t \, < \, 68 \, \rm km \, s^{-1}$, see Ferdman et al. 2014\nocite{fsk+14}).
Furthermore, neither PSR~J0737$-$3039A \citep{fsk+13} nor
PSR~J1751$-$2251 \citep{fsk+14} show any changes in their pulse profiles as a
result of geodetic precession.
The likely reason is the following: after these pulsars were spun up by
material from the companion, their spin angular momentum was
closely aligned with the orbital angular momentum of the system. After the
second SN, its small kick produced only a small change in the orbital plane
of the system; therefore the spin angular momentum of these pulsars
continues to be closely aligned with the orbital angular momentum. In the case of
PSR~J0737$-$3039A, the angle between spin axis and
orbital angular momentum is smaller than $3^\circ$  \citep{fsk+13};
for PSR~J1756$-$2251, this angle is smaller than 17$^\circ$ \citep{fsk+14}.
This implies that, as it precesses around the total angular momentum of the system,
the spin axis of the pulsar will only change direction slightly, resulting
in undetectably small changes in the radio pulse profile as observed from the Earth.

Since the SN kick does not influence the inclination, it also has a limited
influence on the post-SN eccentricity of the orbit. The small orbital eccentricities in
these systems are in large part due to the sudden loss of the NS binding energy
(as neutrinos)
during the second supernova, see e.g. Bhattacharya \& van den Heuvel (1991)\nocite{bh91}.

Like the latter systems, J1913+1102 has a low eccentricity and is also located near
the Galactic plane ($b = 0.19$\deg) despite its large age; both of these are consistent with
a second SN having a small kick. In this case, we should expect, by analogy
with what was said above, that:
a) A small peculiar velocity and resulting small proper motion,
b) No observation of changes in the pulse profile of the recycled (first formed)
pulsar due to geodetic precession, as for PSR~J0737$-$3039A and PSR~J1756$-$2251, and
c) The mass for the NS companion to PSR~J1913+1102 should be similar to
those observed among the second-formed NSs in the  the ``low-kick" systems, i.e.
in the range 1.23\,--\,1.29\,\msun \citep{fsk+14,lks+15}.
All of these characteristics will be determined precisely with continued timing of the
pulsar.

\section{Prospects}
\label{sec:prospects}

If the prediction of a low-mass companion to PSR~J1913+1102 is confirmed, then
the system must be very asymmetric: its total mass would imply $M_p
\,\sim$1.65\,\msun, and thus $q \, \sim\, 1.35$.
This might appear to be unusual, but a similarly asymmetric
DNS system has already been observed, J0453+1559, where
$M_p = 1.559(5)$\,\msun, $M_c = 1.174(4)$\,\msun and $q\, = \, 1.33$  \citep{msf+15}.
As for all the other compact DNSs, continued timing of \psr
should allow the measurement of a second post-Keplerian parameter, the Einstein
delay ($\gamma$), and allow a measurement of the individual NS masses in this
system.

A mass ratio $q$ larger than 1.3 leads to a peculiar behavior during
NS-NS mergers: the lighter (and larger) NS is tidally
disrupted by the smaller, more massive NS. According to recent simulations
\citep{rbg+10,hkk+13,ros13}, such mergers result in a much larger release of
heavy r-process elements \citep{jbp+15}, possibly explaining the heavy
element abundances in our Galaxy. However, until now it was impossible to verify
whether this scenario happens in the Universe because no asymmetric DNSs are
known that might coalesce in a Hubble time. It is therefore very important to
determine whether J1913+1102 is really as asymmetric as the above arguments
suggest.

As observed in the other compact DNS systems, long-term timing will also result
in the precise measurement of another post-Keplerian parameter, the orbital
decay due to the emission of gravitational waves ($\dot{P}_b$). A measurement of
the proper motion will be necessary to ascertain to what degree can we correct
for kinematic effects to the orbital decay of J1913+1102 and test the GR
prediction for $\dot{P}_b$. If the pulsar is co-moving with the local standard 
of rest, we expect a proper motion of about 6 mas/yr. In that case, the
contributions from the Shklovskii effect \citep{Shklovskii70} will
cancel the contribution from differential Galactic acceleration
\citep{dt91,Reid2014} for a wide range of distances around the
estimated DM distance,~7.65 kpc. If the proper motion is much larger then
a precise measurement of the distance will be necessary for 
a precise test of the radiative properties of gravity. Such a precise
distance measurement is unlikely given the low flux density of the pulsar
and the large estimated distance.

As mentioned above, this system will coalesce within 0.5 Gyr. With this
coalescence timescale, it is unlikely that this discovery will have a
large impact on the estimated rate of NS-NS merger events detectable by
ground-based gravitational wave detectors. However, if the mass asymmetry is
confirmed, it will be important for
estimating the rate of events where tidal disruption of a NS is observable.

\section{Conclusions}
\label{sec:conclusion}
We have presented the discovery and timing of \psr, a 27.3-ms pulsar in a binary
orbit with a neutron star companion found in the PALFA survey. Searches of 44
observations for periodic signals from the companion did not result in a
detection. A timing analysis of $\sim1.5$\,yr of data provided a detection of
the rate of advance of the periastron of the orbit, which, assuming GR, resulted
in a measurement of the total binary system mass of $M_\mathrm{tot} =
2.875(14)$\,\msun, making this the most massive DNS system known. A comparison
of J1913+1102 and other DNS systems suggest that its companion may be a low-mass NS
formed via an ultra-stripped SN; in this case the system would be very
asymmetric. This possibility will be tested by the end of 2016 by the
detection of additional PK parameters.

\acknowledgements
We thank all Einstein@Home volunteers, especially those whose computers found
PSR J1913+1102 with the highest statistical significance: Uwe Tittmar,
Kressbronn, Germany, and Gerald Schrader, San Diego, USA.
This work was supported by the Max Plank Society and by NSF grants
1104902, 1105572, and 1148523.

The Arecibo Observatory is operated by SRI International under a cooperative
agreement with the National Science Foundation (AST-1100968), and in alliance
with Ana G. M\'endez-Universidad Metropolitana, and the Universities Space
Research Association. We thank all the dedicated staff that work at
Arecibo, without their efforts and professionalism this discovery would
have been impossible.

Computations were made on the supercomputer Guillimin at McGill University,
managed by Calcul Qu\'{e}bec and Compute Canada. The operation of this
supercomputer is funded by the Canada Foundation for Innovation (CFI),
NanoQu\'{e}bec, RMGA and the Fonds de recherche du Qu\'{e}bec - Nature et
technologies (FRQ-NT).

PL acknowledges the support of IMPRS Bonn/Cologne.
PL, PCCF, and LGS gratefully acknowledge financial support
by the European Research Council for the ERC Starting Grant
BEACON under contract no. 279702.
JSD was supported by the NASA Fermi Guest Investigator program and by the Chief
of Naval Research.
JWTH acknowledges funding from an NWO Vidi fellowship and ERC Starting Grant
``DRAGNET'' (337062).
JvL acknowledges funding from the European Research Council under the European
Union's Seventh Framework Programme (FP/2007-2013) / ERC Grant Agreement n.
617199.
Pulsar research at UBC is supported by an NSERC Discovery Grant and by the
Canadian Institute for Advanced Research.
V.M.K. receives support from an NSERC Discovery Grant and Accelerator Supplement, 
from NSERC's Herzberg Award, from an R. Howard Webster Foundation Fellowship from
the Canadian Institute for Advanced Study, the Canada Research Chairs Program, and the
Lorne Trottier Chair in Astrophysics and Cosmology.

This research has made use of NASA's Astrophysics Data System Bibliographic Services.

We also thank the anonymous referee for suggestions that have improved
this work.

\bibliographystyle{apj}

\end{document}

%% file: githash.tex
Draft version date 2016 August 30 